\documentclass[letterpaper, 10 pt, conference]{ieeeconf}
\IEEEoverridecommandlockouts
\usepackage{amsfonts,amsmath,amssymb} 

\usepackage{psfrag,color}
\usepackage{enumerate,cite,latexsym,graphicx}
\newtheorem{theorem}{Theorem}

\newtheorem{definition}{Definition}

\title{\LARGE \bf Quantum Robust Stability of a Small Josephson Junction in a Resonant Cavity 
}

\author{Ian R.~Petersen  %
\thanks{This work was supported by the
Australian Research Council (ARC) and Air Force Office of Scientific
Research (AFOSR). This material is based on research sponsored by the
Air Force Research Laboratory, under agreement number
FA2386-09-1-4089.  The U.S. Government is authorized to reproduce and
distribute reprints for Governmental purposes notwithstanding any
copyright notation thereon.
The views and conclusions contained herein are those of the authors
and should not be interpreted as necessarily representing the official
policies or endorsements, either expressed or implied, of the Air
Force Research Laboratory or the U.S. Government. }%
\thanks{Ian R. Petersen is with the School of  Engineering and Information Technology, 
        University of New South Wales at the Australian Defence Force Academy, Canberra ACT 2600, Australia.
         {\tt\small i.r.petersen@gmail.com} } 
}%

\begin{document}
\maketitle
\thispagestyle{empty}
\pagestyle{empty}

\begin{abstract}
This paper applies recent results on the robust stability of nonlinear quantum systems to the case of a Josephson junction in a resonant cavity. The Josephson junction is characterized by a Hamiltonian operator which contains a non-quadratic term involving a cosine function. This leads to a sector bounded nonlinearity which enables the previously developed theory to be applied to this system in order to analyze its stability.  
\end{abstract}

\section{Introduction} \label{sec:intro}
In recent years, a number of papers have considered the feedback
control of systems whose dynamics are governed by the laws of quantum
mechanics rather than classical mechanics; e.g., see
\cite{YK03A,YK03B,YAM06,JNP1,NJP1,GGY08,MaP3,MaP4,YNJP1,GJ09,GJN10,WM10,PET10Ba}. In
particular, the papers \cite{GJ09,JG10} consider a framework of
quantum systems defined in terms of a triple $(S,L,H)$ where $S$ is a
scattering matrix, $L$ is a vector of coupling operators and $H$ is a
Hamiltonian operator. 

The paper \cite{PUJ1a} considers the problem of absolute stability for a quantum system defined in terms of a triple $(S,L,H)$ in which the quantum system Hamiltonian is
decomposed as $H =H_1+H_2$ where $H_1$ is a known nominal Hamiltonian
and $H_2$ is a perturbation Hamiltonian, which is contained in a
specified set of Hamiltonians $\mathcal{W}$. In particular the paper \cite{PUJ1a} considers the case in which the
nominal Hamiltonian $H_1$ is a quadratic function of annihilation and
creation operators and the coupling operator vector is a linear
function of annihilation and creation operators. This case corresponds
to a nominal  linear quantum system; e.g., see
\cite{JNP1,NJP1,MaP3,MaP4,PET10Ba}. Also, it is assumed that $H_2$ is contained in a set of non-quadratic perturbation Hamiltonians corresponding to a sector bound on the nonlinearity In this special case, \cite{PUJ1a} obtains a robust stability result in terms of a frequency domain condition. 

In this paper, we apply the result of \cite{PUJ1a} to a quantum system
which consists of a Josephson junction in a resonant cavity as
described in the paper \cite{AS01}. This enables us to analyze the
stability of this quantum system. In particular, this enables us to
choose suitable values for the coupling parameters in the system. For the parameter values chosen, we show that the quantum system is robustly mean square stable according to the definition of \cite{PUJ1a}. 
\section{Quantum Systems} \label{sec:systems}
The main result of the paper \cite{PUJ1a}  considers  open quantum systems defined by  parameters $(S,L,H)$ where $H = H_1+H_2$; e.g., see \cite{GJ09,JG10}.  The corresponding generator for this quantum system is given by 
\begin{equation}
\label{generator}
\mathcal{G}(X) = -i[X,H] + \mathcal{L}(X)
\end{equation}
where $ \mathcal{L}(X) = \frac{1}{2}L^\dagger[X,L]+\frac{1}{2}[L^\dagger,X]L$. Here, $[X,H] = XH-HX$ denotes the commutator between two operators and the notation $^\dagger$ denotes the adjoint transpose of a vector of operators. Also, $H_1$ is a self-adjoint operator on the underlying Hilbert space referred to as the nominal Hamiltonian and $H_2$ is a self-adjoint operator on the underlying Hilbert space referred to as the perturbation Hamiltonian.  The triple $(S,L,H)$, along with the corresponding generators define the Heisenberg evolution $X(t)$ of an operator $X$ according to a quantum stochastic differential equation; e.g., see \cite{JG10}.

We now define a  set of non-quadratic perturbation
Hamiltonians denoted $\mathcal{W}$.   
The set $\mathcal{W}$ is defined in terms of the following  power series (which is assumed to converge in the sense of the induced operator norm on the underlying Hilbert space)
\begin{equation}
\label{H2nonquad}
H_2 = f(\zeta,\zeta^*) = \sum_{k=0}^\infty\sum_{\ell=0}^\infty S_{k\ell}\zeta^k(\zeta^*)^\ell = \sum_{k=0}^\infty \sum_{\ell=0}^\infty S_{k\ell} H_{k\ell}.
\end{equation}
Here $S_{k\ell}=S_{\ell k}^*$, $H_{k\ell} = \zeta^k(\zeta^*)^\ell$, and $\zeta$ is a scalar operator on the underlying Hilbert space. Also, $^*$ denotes the adjoint of a scalar operator. It follows from this definition that
\[
H_2^* = \sum_{k=0}^\infty\sum_{\ell=0}^\infty S_{k\ell}^*\zeta^\ell(\zeta^*)^k = 
\sum_{\ell=0}^\infty\sum_{k=0}^\infty S_{\ell k}\zeta^\ell(\zeta^*)^k = H_2
\]
and thus $H_2$ is a self-adjoint operator. Note that it follows from
the use of Wick ordering that the form (\ref{H2nonquad}) is the most
general form for a perturbation Hamiltonian defined in terms of a
single scalar operator $\zeta$. 

Also, we let 
\begin{equation}
\label{fdash}
f'(\zeta,\zeta^*) = \sum_{k=1}^\infty\sum_{\ell=0}^\infty k S_{k \ell} \zeta^{k-1}(\zeta^*)^\ell,
\end{equation}
\begin{equation}
\label{fddash}
f''(\zeta,\zeta^*) = \sum_{k=1}^\infty\sum_{\ell=0}^\infty k(k-1)S_{k\ell} \zeta^{k-2}(\zeta^*)^{\ell}
\end{equation}
and consider the 
sector bound condition
\begin{equation}
\label{sector4a}
f'(\zeta,\zeta^*)^*f'(\zeta,\zeta^*)  \leq \frac{1}{\gamma^2}\zeta \zeta^* + \delta_1
\end{equation}
and the condition
\begin{equation}
\label{sector4b}
f''(\zeta,\zeta^*)^*f''(\zeta,\zeta^*) \leq  \delta_2.
\end{equation}
Then we define the set $\mathcal{W}$  as follows:
\begin{equation}
\label{W5}
\mathcal{W} = \left\{\begin{array}{l}H_2 \mbox{ of the form
      (\ref{H2nonquad}) such that 
} \\
\mbox{ conditions (\ref{sector4a}) and (\ref{sector4b}) are satisfied}\end{array}\right\}.
\end{equation}

Reference \cite{PUJ1a} also  considers the  case in which the nominal quantum system corresponds to a linear quantum system; e.g., see \cite{JNP1,NJP1,MaP3,MaP4,PET10Ba}. In this case, $H_1$ is of the form 
\begin{equation}
\label{H1}
H_1 = \frac{1}{2}\left[\begin{array}{cc}a^\dagger &
      a^T\end{array}\right]M
\left[\begin{array}{c}a \\ a^\#\end{array}\right]
\end{equation}
where $M \in \mathbb{C}^{2n\times 2n}$ is a Hermitian matrix of the
form
\begin{equation}
\label{Mform}
M= \left[\begin{array}{cc}M_1 & M_2\\
M_2^\# &     M_1^\#\end{array}\right]
\end{equation}
and $M_1 = M_1^\dagger$, $M_2 = M_2^T$.  Here, the notation $^\#$ denotes the vector of adjoint operators for a vector of operators. Also, $^\#$ denotes denotes the complex conjugate of a matrix for a complex matrix. In addition, we assume $L$ is of the form 
\begin{equation}
\label{L}
L = \left[\begin{array}{cc}N_1 & N_2 \end{array}\right]
\left[\begin{array}{c}a \\ a^\#\end{array}\right]
\end{equation}
where $N_1 \in \mathbb{C}^{m\times n}$ and $N_2 \in
\mathbb{C}^{m\times n}$. Also, we write
\[
\left[\begin{array}{c}L \\ L^\#\end{array}\right] = N
\left[\begin{array}{c}a \\ a^\#\end{array}\right] =
\left[\begin{array}{cc}N_1 & N_2\\
N_2^\# &     N_1^\#\end{array}\right]
\left[\begin{array}{c}a \\ a^\#\end{array}\right].
\]

As in \cite{PUJ1a}, we consider a notion of robust mean square stability. 
\begin{definition}
\label{D1}
An uncertain open quantum system defined by  $(S,L,H)$ where $H=H_1+H_2$ with $H_1$ of the form (\ref{H1}), $H_2 \in \mathcal{W}$, and $L$  of the form (\ref{L}) is said to be {\em robustly mean square stable} if for any $H_2 \in \mathcal{W}$, there exist constants $c_1 > 0$, $c_2 > 0$ and $c_3 \geq 0$ such that
\begin{eqnarray}
\label{ms_stable0}
\lefteqn{\left< \left[\begin{array}{c}a(t) \\ a^\#(t)\end{array}\right]^\dagger \left[\begin{array}{c}a(t) \\ a^\#(t)\end{array}\right] \right>}\nonumber \\
&\leq& c_1e^{-c_2t}\left< \left[\begin{array}{c}a \\ a^\#\end{array}\right]^\dagger \left[\begin{array}{c}a \\ a^\#\end{array}\right] \right>
+ c_3~~\forall t \geq 0.
\end{eqnarray}
Here $\left[\begin{array}{c}a(t) \\ a^\#(t)\end{array}\right]$ denotes the Heisenberg evolution of the vector of operators $\left[\begin{array}{c}a \\ a^\#\end{array}\right]$; e.g., see \cite{JG10}.
\end{definition}

We define
\begin{eqnarray}
\label{z}
\zeta &=&  E_1a+E_2 a^\# \nonumber \\
&=& \left[\begin{array}{cc} E_1 & E_2 \end{array}\right]
\left[\begin{array}{c}a \\ a^\#\end{array}\right] = \tilde E 
\left[\begin{array}{c}a \\ a^\#\end{array}\right]
\end{eqnarray}
where $\zeta$ is  a scalar operator. Then,  the following 
 following strict bounded real condition
provides a sufficient condition for the robust mean square stability
of the nonlinear quantum system under consideration when $H_2 \in \mathcal{W}$: 
\begin{enumerate}
\item
The matrix 
\begin{equation}
\label{Hurwitz1}
F = -iJM-\frac{1}{2}JN^\dagger J N\mbox{ is Hurwitz;}
\end{equation}
\item
\begin{equation}
\label{Hinfbound1}
\left\|\tilde E^\# \Sigma\left(sI -F\right)^{-1}\tilde D \right\|_\infty < \frac{\gamma}{2}
\end{equation}
where 
\begin{eqnarray*}
\tilde D &=& J\Sigma \tilde E^T, \\
J&=& \left[\begin{array}{cc} I & 0\\
0 &-I \end{array}\right],\\
\Sigma &=& \left[\begin{array}{cc} 0 & I\\
I &0 \end{array}\right].
\end{eqnarray*}
\end{enumerate}

This leads to the following theorem which is presented in \cite{PUJ1a}.

\begin{theorem}
\label{T4}
Consider an uncertain open quantum system defined by $(S,L,H)$  such that
$H=H_1+H_2$ where $H_1$ is of the form (\ref{H1}), $L$ is of the
form (\ref{L}) and $H_2 \in \mathcal{W}$. Furthermore, assume that
the strict bounded real condition  (\ref{Hurwitz1}), (\ref{Hinfbound1})
is satisfied. Then the
uncertain quantum system is robustly mean square stable. 
\end{theorem}
In the next section, we will apply this theorem to analyze the
stability of a nonlinear quantum system corresponding to a Josephson
junction in a resonant cavity.  
\section{The Josephson Junction in a Resonant Cavity System}
\label{sec:josephson}
We consider a quantum system consisting of a small Josephson junction
coupled to an electromagnetic resonant cavity. This system has been considered in the
paper \cite{AS01} where a Hamiltonian for the system is derived. The
paper \cite{AS01} considers the system as a closed quantum system
defined purely in terms of a system Hamiltonian. We modify this
description of the system to consider an open quantum system model
which interacts with external fields by introducing coupling operators
for the system in order to apply the results of \cite{PUJ1a} given
above. The coupling operator which is introduced is taken as a
standard coupling operator for a resonant cavity coupled to a single
field as well as a corresponding coupling operator for the Josephson
junction coupled to an external heat bath. 

\begin{figure}[hbp]
\begin{center}
\includegraphics[width=8cm]{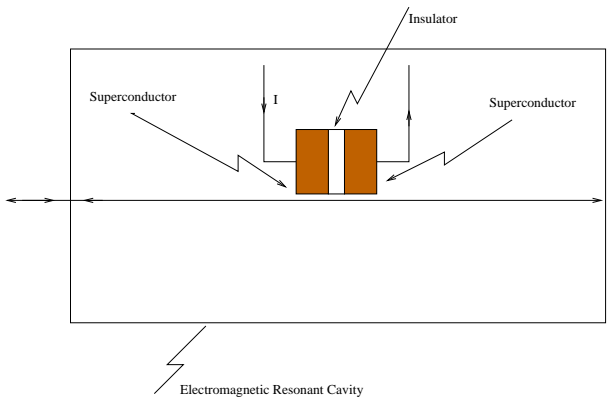}
\end{center}
\caption{Schematic diagram of a Josephson junction in a resonant cavity.}
\label{F1}
\end{figure}

A Josephson junction consists of a thin insulating material between two superconducting layers. As in \cite{AS01}, we consider a Josephson junction in a resonant cavity. This is illustrated in Figure \ref{F1}. 
The following Hamiltonian for the Josephson junction system is derived in \cite{AS01}
\begin{eqnarray}
\label{Hamil}
\mathcal{H} &=& \frac{1}{2\hbar}U'(n'-\bar n')^2 - \frac{J'}{\hbar} \cos \phi' 
+\frac{1}{2\hbar}(p'^2+\omega^2q'^2)\nonumber \\
&&-
g\sqrt{\frac{\omega}{\hbar}}p'n'
+\frac{U \omega \bar n^2 g^2}{2U'}
\end{eqnarray}
where $\bar n' = \frac{U\bar n}{U'}$, $U'=U+\hbar \omega g^2$, $[\phi',n'] = i$, and $[q',p'] = i\hbar$.
Here
\begin{eqnarray*}
\phi'&=&\phi-g\sqrt{\omega/\hbar}q,\quad n'=n\\
p'&=&p+g\sqrt{\omega\hbar}n,\quad q'=q
\end{eqnarray*}
where
$q$ and $p$ are the position and momentum operators for the resonant cavity.
Also, 
$n$ is an operator
which represents the difference between the number of Cooper
pairs on the two superconducting islands which make up
the junction. Furthermore, 
$\phi$ is an operator which represents  the phase difference
across the junction. Note that compared to the expression for the Hamiltonian given in
\cite{AS01}, we have normalized the expression (\ref{Hamil}) by dividing through by a factor of $\hbar$ in order
to be consistent with the convention used in 
\cite{PUJ1a}. 

The quantities $U$, $J'$, $\bar n$, $g$, $\omega$ are physical
constants associated with the junction and the resonant cavity. In
particular, $U$ is the charging energy of the Josephson junction, $J'$
is the Josephson energy of the junction, $\bar n$ is an experimental
parameter relating to the gate voltage applied to the superconducting
islands, $g$ is a parameter related to the dimensions of the junction,
and $\omega$ is an angular frequency related to the detuning of the cavity; see
\cite{AS01}.

By a process of completion of squares, defining new variables $n'' = n'-\bar n$, $p''=p'-g\sqrt{\hbar \omega}$ and neglecting the constant terms, we can re-write the Hamiltonian as follows:
\begin{eqnarray*}
\lefteqn{\mathcal{H}' =}\\
&& \frac{1}{2}[q'~p''~n''~\phi']
\left[\begin{array}{cccc}
\frac{\omega^2}{\hbar} & 0 & 0 & 0\\
0 & \frac{1}{\hbar} & -g\sqrt{ \frac{\omega}{\hbar}} & 0\\
0 & -g\sqrt{ \frac{\omega}{\hbar}} & \frac{U'}{\hbar} & 0\\
0 & 0 & 0 & 0
\end{array}\right]
\left[\begin{array}{c}
q'\\p''\\n''\\\phi'
\end{array}\right]\\
&&-\frac{J'}{\hbar}\cos\phi'.
\end{eqnarray*}
Now we define new operators $a_1 = (\omega q'+ip'')/\sqrt{2\hbar
  \omega}$, and  $a_2 = (\phi'+i n'')/\sqrt{2}$ which satisfy the
canonical commutation relations $[a_1,a_1^*] = 1$ and $[a_2,a_2^*]=1$. 
Then, the Hamiltonian can be re-written in the form
\begin{equation}
\label{Hjosephson}
H = \frac{1}{2}\left[\begin{array}{cc}a^\dagger &
      a^T\end{array}\right]M
\left[\begin{array}{c}a \\ a^\#\end{array}\right] - \frac{J'}{\hbar}\cos(\frac{a_2+a_2^*}{\sqrt2})
\end{equation}
where $a =\left[\begin{array}{c}a_1\\a_2\end{array}\right]$ and $M$ is
a Hermitian matrix of the form (\ref{Mform}). Note that in order to
write the Hamiltonian in this form with the matrix $M$ satisfying
(\ref{Mform}), it is necessary to apply these canonical commutation relations and
neglect further constant terms. Also, the fact that the operators
$a_1$ and $a_2$ commute is used to re-distribute terms within the
matrix $M$. (Formally $a_1$ and $a_2$ are defined on different Hilbert spaces but following the standard convention in quantum mechanics we can extend each operator to the tensor product of the two Hilbert spaces and then these two extended operators will commute with each other.)

We now modify the model of \cite{AS01} by assuming that the cavity and Josephson modes are coupled to  fields corresponding to coupling operators of the form
\[
L = \left[\begin{array}{c}
\sqrt{\kappa_1}a_1\\ \sqrt{\kappa_2}a_2
\end{array}\right].
\]
This modification of the quantum system is necessary in order to
obtain a damped quantum system whose stability can be established
using the approach of \cite{PUJ1a} and is quite reasonable for an
experimental system which will experience damping in both the
electromagnetic cavity and in the Josephson junction circuit. 

In order to apply Theorem \ref{T4} to analyze the stability of this
quantum system, we rewrite (\ref{Hjosephson}) as
\[
H= H_1 +H_2
\]
where $H_1 = \frac{1}{2}\left[\begin{array}{cc}a^\dagger &
      a^T\end{array}\right]M
\left[\begin{array}{c}a \\ a^\#\end{array}\right]$ and $H_2 = -
\frac{J'}{\hbar}\cos(\frac{a_2+a_2^*}{\sqrt2})$. That is, we have a
quadratic nominal Hamiltonian and a non-quadratic perturbation
Hamiltonian. Then, we define $\zeta = a_2/\sqrt2$ and
\begin{eqnarray*}
f(\zeta,\zeta^*) &=& -\frac{J'}{\hbar}\cos(\zeta+\zeta^*)\\
f'(\zeta,\zeta^*) &=& \frac{J'}{\hbar}\sin(\zeta+\zeta^*)\\
f''(\zeta,\zeta^*) &=& \frac{J'}{\hbar}\cos(\zeta+\zeta^*).
\end{eqnarray*}
From this it follows that 
\[
f'(\zeta,\zeta^*)^*f'(\zeta,\zeta^*) \leq 4 \frac{J'^2}{\hbar^2}\zeta \zeta^*, \quad f''(\zeta,\zeta^*)^*f''(\zeta,\zeta^*) \leq \frac{J'^2}{\hbar^2}
\]
and $\gamma = \frac{\hbar}{2J'}$.

The numerical values of the constants $\omega$, $g$,  $U$, and $J'$  are chosen
as in \cite{AS01} as 
$\frac{\omega}{2\pi}=100\mbox{GHz}$, $g=0.15$,  $U=2.2087\times 10^{-22}$, 
$J'=3.6652\times 10^{11}$. Also, we calculate $\gamma/2 = 6.8209\times
10^{-13}$. The parameters  $\kappa_1$ and $\kappa_2$
will be chosen using Theorem \ref{T4} in order guarantee the robust
mean square stability of the quantum system. Indeed, for various
values of $\kappa_1$ and $\kappa_2$, we form the transfer function
$G_{\kappa_1, \kappa_2}(s) = E^\# \Sigma\left(sI -F\right)^{-1}\tilde D$
and calculate its $H_\infty$ norm. 

\begin{figure}[htbp]
\begin{center}
\psfrag{normG}{$\|G_{\kappa_1,\kappa_2}(s)\|_\infty$}
\psfrag{K1}{$\kappa_2$}
\psfrag{gam}{$\frac{\gamma}{2}$}
\includegraphics[width=8cm]{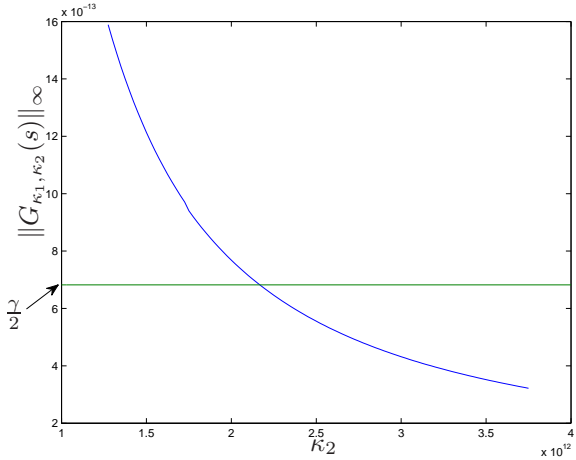}
\end{center}
\caption{Plot of $\|G_{\kappa_1, \kappa_2}(s)\|_\infty$ versus $\kappa_2$.}
\label{F2}
\end{figure}

The $H_\infty$ norm of $G_{\kappa_1, \kappa_2}(s)$ was found to
be virtually independent of $\kappa_1$ and so a physically reasonable
value of $\kappa_1=10^{11}$ was chosen. With this value of $\kappa_1$,
a plot of $\|G_{\kappa_1, \kappa_2}(s)\|_\infty$ versus $\kappa_2$ is
shown in Figure \ref{F2}.

\begin{figure}[htbp]
\begin{center}
\includegraphics[width=8cm]{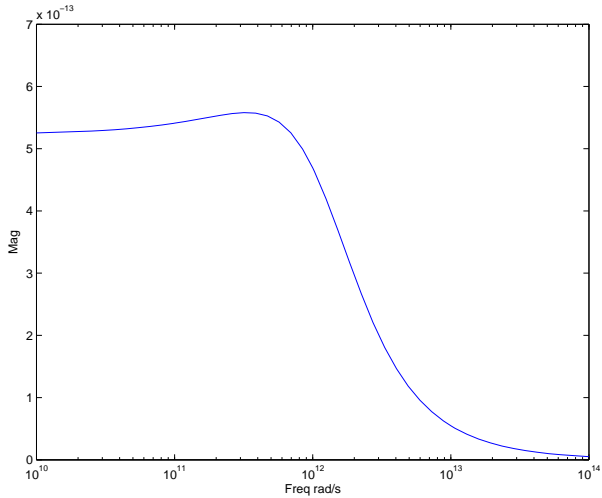}
\end{center}
\caption{Magnitude Bode plot of $G_{\kappa_1, \kappa_2}(s)$.}
\label{F3}
\end{figure}

From this plot we can see that stability can be guaranteed for $\kappa_2 > 2.2\times 10^{12}$. Hence, choosing a value of $\kappa_2 = 2.5\times 10^{12}$, it follows that stability of Josephson junction system can be guaranteed using Theorem \ref{T4}. Indeed, with this value of $\kappa_2$, we calculate the matrix $F = -iJM-\frac{1}{2}JN^\dagger J N$ and find its eigenvalues to be   
$-5.0000\times 10^{10} \pm 3.3507\times 10^3i$ and 
$ -1.2500\times 10^{12} \pm 1.4842\times 10^3i$ which implies that the matrix $F$ is Hurwitz. 
Also, a magnitude Bode plot of the corresponding transfer function
$G_{\kappa_1, \kappa_2}(s)$ is shown in Figure \ref{F3} below which
implies that $\|G_{\kappa_1, \kappa_2}(s)\|_\infty = 5.5554\times
10^{-13} < \gamma/2 = 6.8209\times10^{-13}$. Hence, using Theorem
\ref{T4}, we conclude that the quantum system is robustly mean square stable. 

\section{Conclusions}
\label{sec:conclusions}
We have applied a previous result on the robust stability of nonlinear
quantum systems to a quantum system arising from a Josephson junction
coupled to an electromagnetic resonant cavity. This system involved a
cosine function in the system Hamiltonian which was a suitable
non-quadratic Hamiltonian leading to a quantum system with a sector
bounded nonlinearity. For specific numerical values of the system
parameters, it was shown that the robust stability result can be used
to choose  coupling parameters for the Josephson junction system in order to
guarantee robust mean square stability.


\begin{thebibliography}{10}
\providecommand{\url}[1]{#1}
\csname url@rmstyle\endcsname
\providecommand{\newblock}{\relax}
\providecommand{\bibinfo}[2]{#2}
\providecommand\BIBentrySTDinterwordspacing{\spaceskip=0pt\relax}
\providecommand\BIBentryALTinterwordstretchfactor{4}
\providecommand\BIBentryALTinterwordspacing{\spaceskip=\fontdimen2\font plus
\BIBentryALTinterwordstretchfactor\fontdimen3\font minus
  \fontdimen4\font\relax}
\providecommand\BIBforeignlanguage[2]{{%
\expandafter\ifx\csname l@#1\endcsname\relax
\typeout{** WARNING: IEEEtran.bst: No hyphenation pattern has been}%
\typeout{** loaded for the language `#1'. Using the pattern for}%
\typeout{** the default language instead.}%
\else
\language=\csname l@#1\endcsname
\fi
#2}}

\bibitem{YK03A}
M.~Yanagisawa and H.~Kimura, ``Transfer function approach to quantum
  control-part {I}: Dynamics of quantum feedback systems,'' \emph{IEEE
  Transactions on Automatic Control}, vol.~48, no.~12, pp. 2107--2120, 2003.

\bibitem{YK03B}
------, ``Transfer function approach to quantum control-part {II}: Control
  concepts and applications,'' \emph{IEEE Transactions on Automatic Control},
  vol.~48, no.~12, pp. 2121--2132, 2003.

\bibitem{YAM06}
N.~Yamamoto, ``Robust observer for uncertain linear quantum systems,''
  \emph{Phys. Rev. A}, vol.~74, pp. 032\,107--1 -- 032\,107--10, 2006.

\bibitem{JNP1}
M.~R. James, H.~I. Nurdin, and I.~R. Petersen, ``${H}^\infty$ control of linear
  quantum stochastic systems,'' \emph{IEEE Transactions on Automatic Control},
  vol.~53, no.~8, pp. 1787--1803, 2008.

\bibitem{NJP1}
H.~I. Nurdin, M.~R. James, and I.~R. Petersen, ``Coherent quantum {LQG}
  control,'' \emph{Automatica}, vol.~45, no.~8, pp. 1837--1846, 2009.

\bibitem{GGY08}
J.~Gough, R.~Gohm, and M.~Yanagisawa, ``Linear quantum feedback networks,''
  \emph{Physical Review A}, vol.~78, p. 062104, 2008.

\bibitem{MaP3}
A.~I. Maalouf and I.~R. Petersen, ``Bounded real properties for a class of
  linear complex quantum systems,'' \emph{IEEE Transactions on Automatic
  Control}, vol.~56, no.~4, pp. 786 -- 801, 2011.

\bibitem{MaP4}
------, ``Coherent ${H}^{\infty}$ control for a class of linear complex quantum
  systems,'' \emph{IEEE Transactions on Automatic Control}, vol.~56, no.~2, pp.
  309--319, 2011.

\bibitem{YNJP1}
N.~Yamamoto, H.~I. Nurdin, M.~R. James, and I.~R. Petersen, ``Avoiding
  entanglement sudden-death via feedback control in a quantum network,''
  \emph{Physical Review A}, vol.~78, no.~4, p. 042339, 2008.

\bibitem{GJ09}
J.~Gough and M.~R. James, ``The series product and its application to quantum
  feedforward and feedback networks,'' \emph{IEEE Transactions on Automatic
  Control}, vol.~54, no.~11, pp. 2530--2544, 2009.

\bibitem{GJN10}
J.~E. Gough, M.~R. James, and H.~I. Nurdin, ``Squeezing components in linear
  quantum feedback networks,'' \emph{Physical Review A}, vol.~81, p. 023804,
  2010.

\bibitem{WM10}
H.~M. Wiseman and G.~J. Milburn, \emph{Quantum Measurement and Control}.\hskip
  1em plus 0.5em minus 0.4em\relax Cambridge University Press, 2010.

\bibitem{PET10Ba}
I.~R. Petersen, ``Quantum linear systems theory,'' in \emph{Proceedings of the
  19th International Symposium on Mathematical Theory of Networks and Systems},
  Budapest, Hungary, July 2010.

\bibitem{JG10}
M.~James and J.~Gough, ``Quantum dissipative systems and feedback control
  design by interconnection,'' \emph{IEEE Transactions on Automatic Control},
  vol.~55, no.~8, pp. 1806 --1821, August 2010.

\bibitem{PUJ1a}
I.~R. Petersen, V.~Ugrinovskii, and M.~R. James, ``Robust stability of
  uncertain quantum systems,'' in \emph{Proceedings of the 2012 American
  Control Conference}, Montreal, Canada, June 2012.

\bibitem{AS01}
W.~Al-Saidi and D.~Stroud, ``Eigenstates of a small {J}osephson junction
  coupled to a resonant cavity,'' \emph{Physical Review B}, vol.~65, p. 014512,
  2001.

\end{thebibliography}

\end{document}